\documentclass{PoS}
\usepackage{url}

\title{The MUonE experiment: a novel way to measure the leading order hadronic contribution to the muon g-2}

\ShortTitle{The MUonE experiment}

\author{\speaker{Graziano Venanzoni}
  \thanks{For the MUonE Collaboration}\\
       INFN Sezione di Pisa, Pisa, Italy\\
        E-mail: \email{graziano.venanzoni@pi.infn.it}}

\abstract{We present the status of the MUonE experimental proposal which aims
  at determining the leading order hadronic contribution to the muon g-2 by measuring the hadronic part of the photon vacuum polarization in the spacelike region. The challenges posed by this measurement on the detector, the proposed solution, and the status of this proposal will be discussed.}

\FullConference{The 39th International Conference on High Energy Physics (ICHEP2018)\\
		4-11 July, 2018\\
		Seoul, Korea}

\begin{document}

\section{Introduction}
\begin{figure*}[htp]
\includegraphics[width=.5\textwidth]{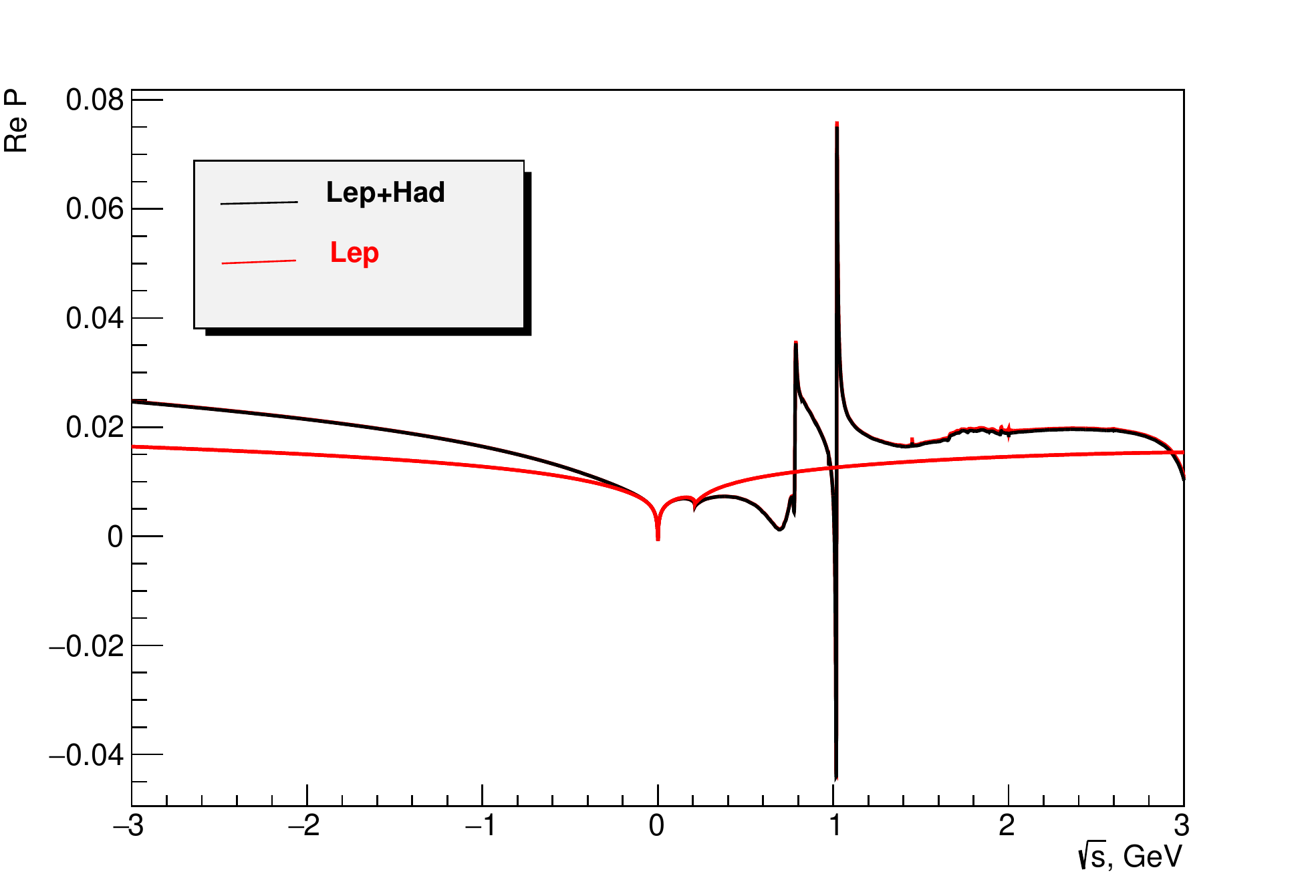}~\includegraphics[width=.5\textwidth]{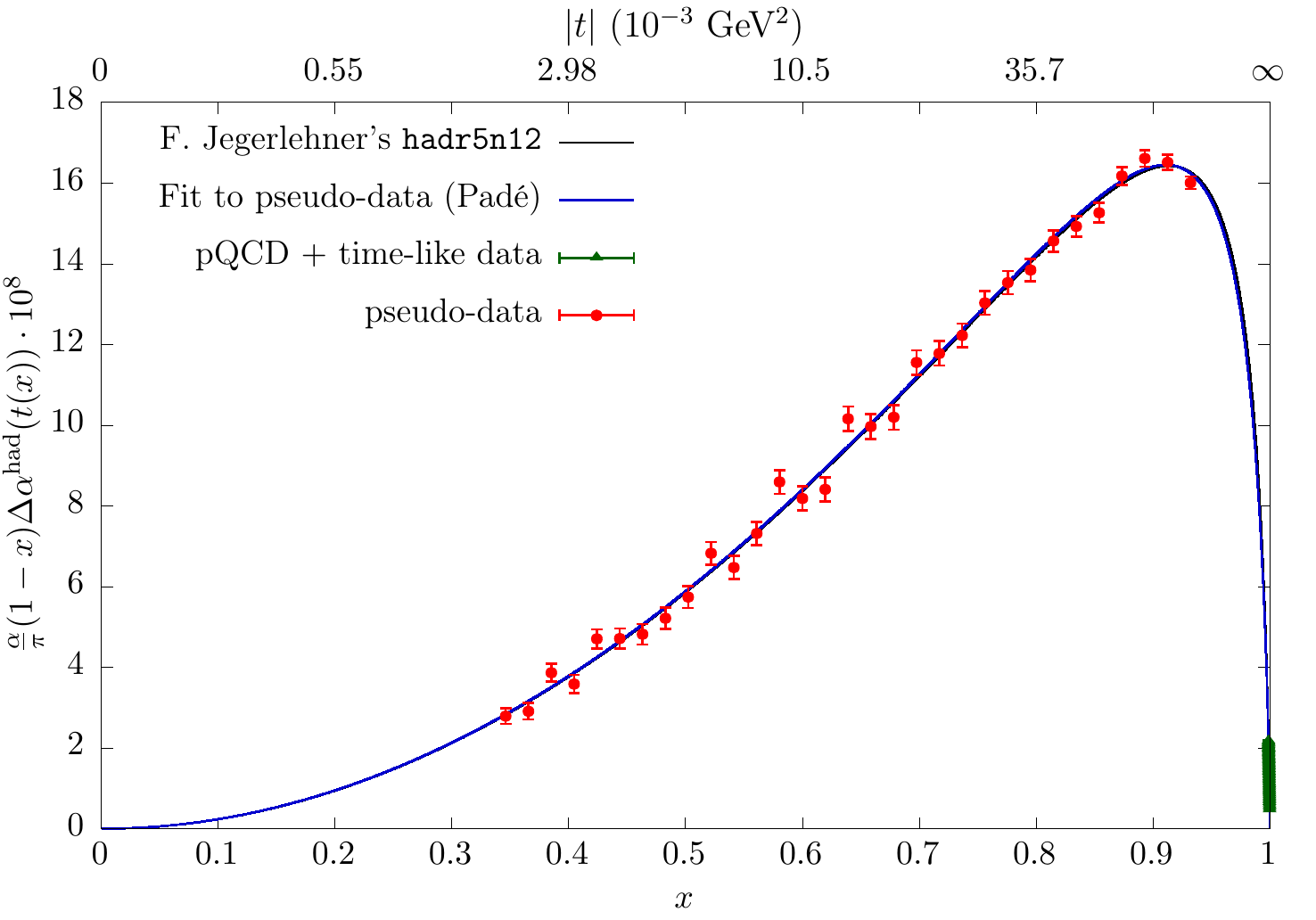}
\caption{{\bf Left:} Real part of the photon vacuum polarization ($\Delta\alpha(q^2))$ in the spacelike and timelike region (Black line: leptonic plus
hadronic contributions; Red line: only leptonic contribution); {\bf Right:}  The integrand $(1-x)\Delta\alpha_{\rm had}[t(x)] \times 10^5$
    as a function of $x$ and $t$ $(t(x)=\frac{x^2m^2_{\mu}}{x-1}<0)$. The peak value is at 
    $x_{\rm peak}\simeq 0.914$, corresponding to $t_{\rm peak} \simeq-0.108$ GeV$^2$. The points refer to an integrated luminosity \mbox{$L = 1.5 \times 10^{7}$~nb$^{-1}$}, achievable at CERN M2 beamline in 2 years of data taking.}
  \label{fig1}
\end{figure*}

There is a tantalizing discrepancy of $\sim 3.5$ standard deviations between the measurement of the muon anomaly $a_{\mu}=(g-2)/2$ performed by the E821 experiment at BNL and the Standard Model prediction~\cite{Blum:2013xva}. Whether this discrepancy is real or not, it certainly calls for a more precise determination of $a_{\mu}$.
New experiments at Fermilab (E989, an evolution of E821) and at J-PARC (E34, with a completely different technique) aim to measure $a_{\mu}$ to 0.14 ppm.
With the planned improvement of the measurement, it's important that the theoretical prediction improves as well.
The leading-order hadronic vacuum polarization contribution, $a_{\mu}^{\rm{HLO}}$,
currently represents the main limitation for the theory due to the non-perturbative QCD behavior at low energy. An intense research program is underway with both timelike data and lattice calculations~\cite{TH}. Recently a new method
to calculate $a_{\mu}^{\rm{HLO}}$ with experimental input from direct measurement of the hadronic part of the photon vacuum polarization ($\Delta\alpha^{had}(q^2)$) in the spacelike region has been proposed~\cite{Calame:2015fva}.
As it can be seen in Fig.~\ref{fig1}, Left\footnote{The plot has been produced with the code
  available at Fedor Ignatov's web page {\url {http://cmd.inp.nsk.su/~ignatov/vpl/.}}},
  the real part of the photon vacuum polarization is a smooth function in the spacelike domain ({\it i.e.} at negative squared momentum transfer), contrary to the timelike behavior where it undergoes significant variations due to the presence of resonances and threshold effects.

\section{The MUonE experiment}
The MUonE experiment, recently proposed at CERN, plans to measure
the hadronic part of the
running of the electromagnetic coupling constant in the spacelike
region by scattering high-energy muons on atomic electrons
of a low-Z target through the elastic process $\mu e\to \mu e$~\cite{Abbiendi:2016xup}.
The differential cross section of this process, measured as a
function of the squared momentum transfer $t =q^2 < 0$,
provides direct sensitivity to the leading-order hadronic contribution
to the muon anomaly, $a_{\mu}^{HLO}$~\cite{Calame:2015fva}.
Assuming a 150~GeV muon beam with an average intensity of $\sim 1.3 \times 10^7$ muons/s, presently available at the CERN muon M2 beamline, incident on a target consisting of 60 Beryllium layers, each 1~cm thick (see Sect. 3), and two years of data taking with a running time of $2 \times 10^7$~s/yr, 
one can reach an integrated luminosity of about $1.5 \times 10^7~{\rm nb}^{-1}$, which would correspond to a statistical error of $0.3\%$ on the value of $a_\mu^{\rm HLO}$, as shown in Fig.~\ref{fig1},
%
 Right, obtained with a simulation of the lowest-order  $\mu e \to \mu e$ cross section. 
\section{Detector considerations and systematic studies}
%



\begin{figure*}[htp]
\includegraphics[width=.5\textwidth]{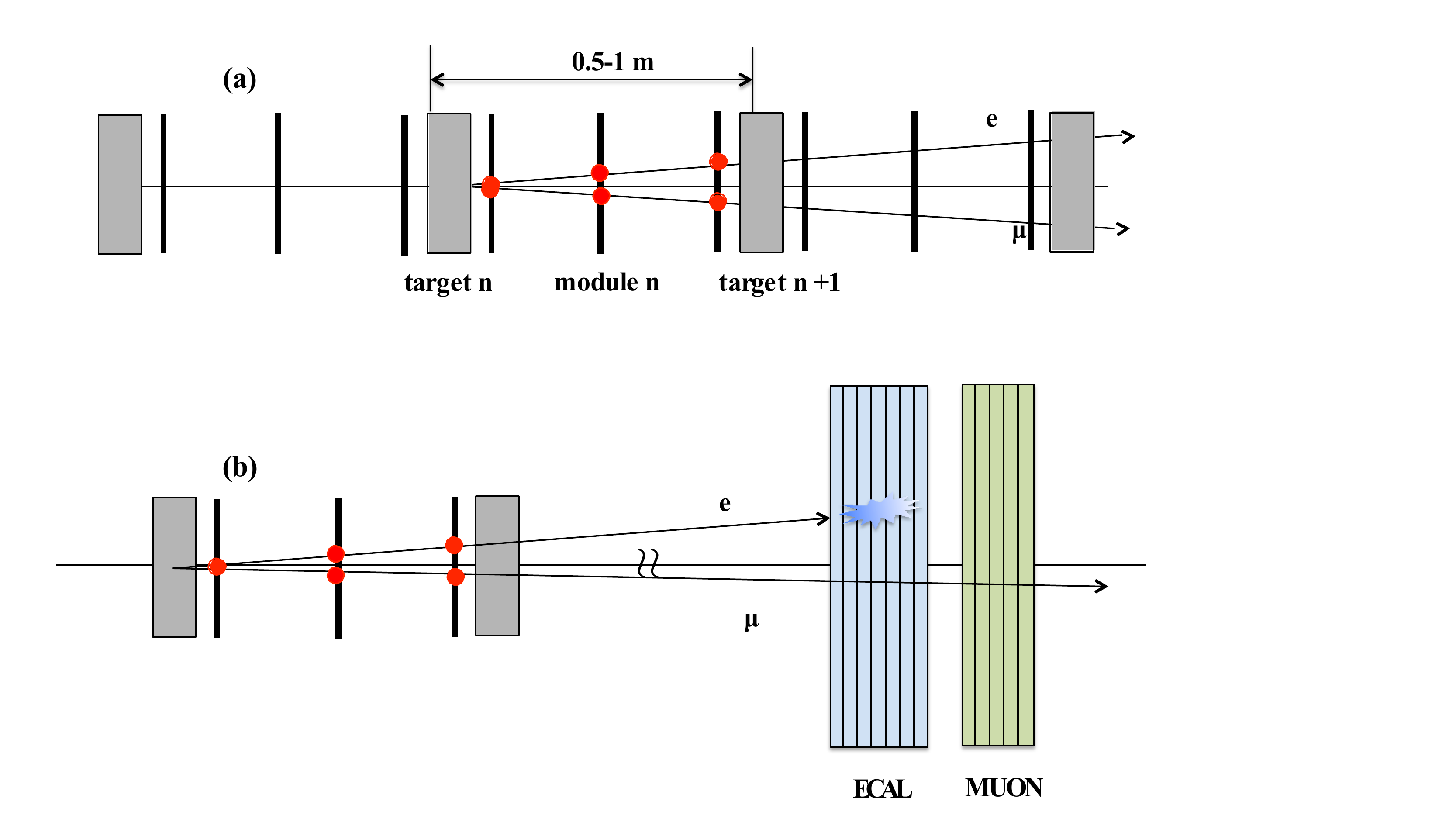}~\includegraphics[width=.5\textwidth]{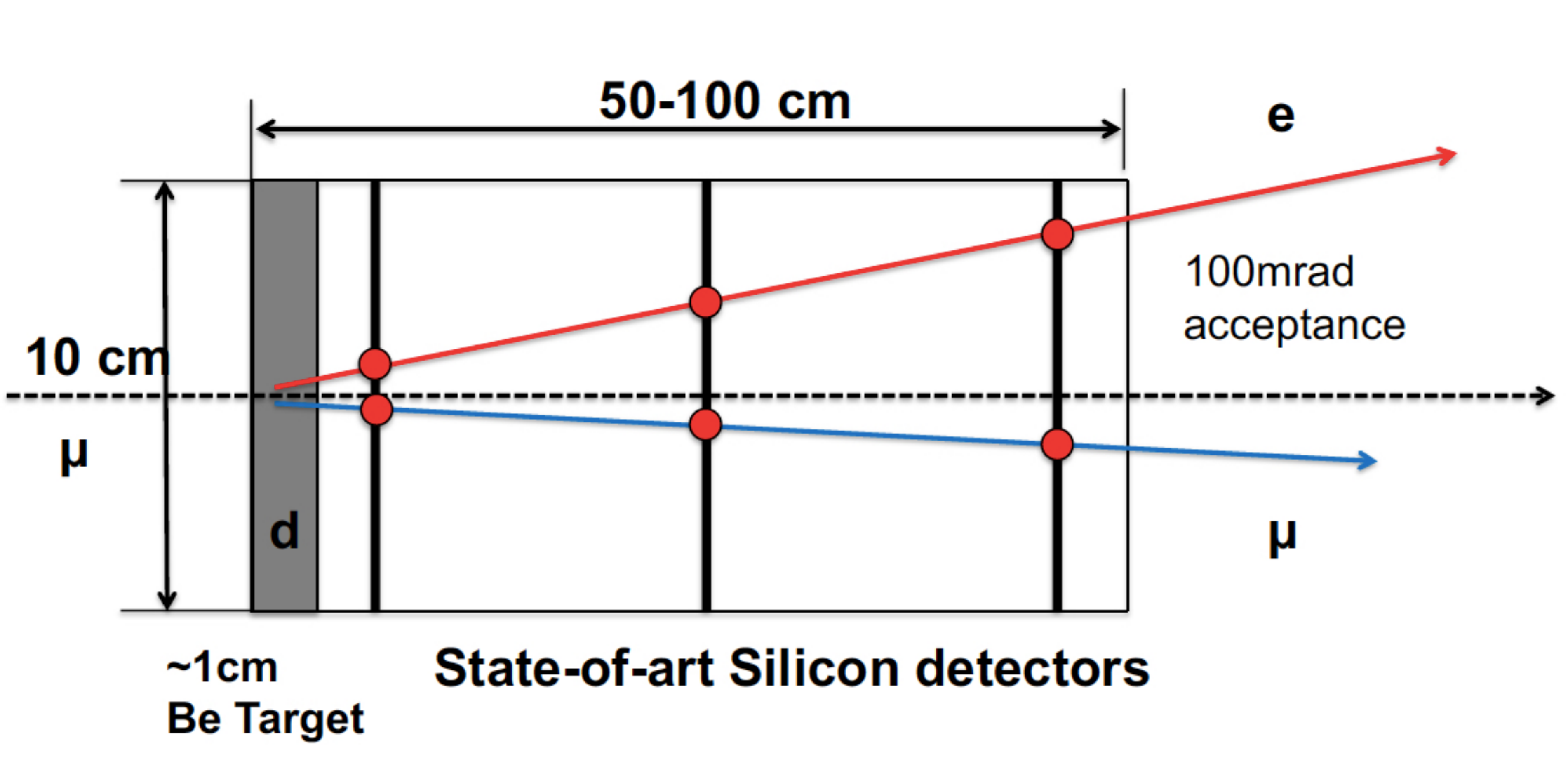}
\caption{{\bf Left:} Design of the baseline detector concept; {\bf Right:} Single unit.}
  \label{Be-detector}
\end{figure*}
Figure~\ref{Be-detector} shows the baseline detector design.
The detector is a repetition of 60 identical modules (called units), 
each consisting of a 1~cm thick layer of Be coupled to 3 Silicon tracking stations 
within a  distance of 50-100 cm (to be optimized) from each other 
with intermediate air gaps.  The choice of thin targets is due to minimize the impact of multiple scattering and the background on the measurement. The number of targets provides the necessary statistics. The Si detectors provide the necessary resolution ($\sim$20 $\mu m$) with a price of a limited material budget ($<$ 0.05 $X_0$ per unit).
The arrangement provides both a distributed target with low-{\it Z} and the tracking system. 
Downstream of the apparatus a calorimeter and a muon system (a filter plus active planes) will be used for e/$\mu$ PID. 
Significant contributions of the hadronic vacuum polarization to the $\mu e \to \mu e$ differential cross section are essentially restricted to electron scattering angles below 10 mrad, corresponding to electron energies above 10 GeV.
The net effect of these contributions is to increase the cross section by a few
 per mille: a precise determination of $a_{\mu}^{\rm{HLO}}$ requires not only
high statistics, but also a high systematic accuracy, as the final goal of the
experiment is equivalent to a determination of the signal to normalization ratio
with $\sim$10 ppm systematic uncertainty at the peak of the integrand function ($cf$. Fig.~\ref{fig1}, Right).
Although this is not a request on the knowledge of the absolute cross section
(signal and normalization regions will be obtained by $\mu e$ data itself)
it poses severe requests on the knowledge of the main following quantities:
\begin{itemize}
\item {Multiple scattering:}
  Preliminary studies indicate that an accuracy of the order of $\sim$ 1\% is required on the knowledge of the Multiple Scattering effects in the core region.
  Results from a Test Beam at CERN with electrons of 12 and 20 GeV on 8-20 mm C target show good agreement between data and GEANT4 simulation.
\item{Tracking uniformity, alignment and reconstruction of angles:}
It's important to keep the systematic error arising from the non-uniformity of the tracking efficiency and angle reconstruction at the 10$^{-5}$ level.
The use of state-of-the-art silicon detectors is sufficient to ensure the required uniformity. The relative alignment of the silicon detectors will be monitored with the high statistics provided by the muon beam; absolute calibration on the transverse plane can be achieved taking advantage of the constrained kinematic of two-body elastic scattering. The longitudinal position of the silicon detector must be monitored in real time at the 10 microns level; this precision can be achieved with commercial laser systems based on interferometry.
\item{Knowledge of the Beam:} 
  A 0.8\% percent accuracy on the knowledge of the beam momentum, as obtained by the BMS spectrometer used by COMPASS, is sufficient to 
control the systematic effects arising from beam spread.

\item{Theory:}
A complete fixed order calculation of QED NNLO radiative corrections, consistently matched with the resummation of higher-orders and implemented into a Monte Carlo event generator, will be required to control the theoretical systematics 
at the 10$^{-5}$ level and, in turn, to extract the leading order hadronic contribution to the muon g-2 with the aimed accuracy. A first NLO MC has been developed by the authors of Babayaga~\cite{Alacevich:2018vez} and NNLO calculations are in progress~\cite{nnloMP}.
\end{itemize}

\section{Current status and future plans of the proposal}
The current Collaboration comprises of groups from Italy, Poland, Russia, UK, and USA. Many of them are experts in the field of precision physics.
Years 2018-2019 will be devoted to detector optimization studies, simulation, test beams and theory improvement with a final goal to present a  Letter of Intent to the SPSC.
The detector construction is expected during LS2 and the plan is to install 
the detector (or a staged version of it)
and start data taking in the period 2021-2024. MUonE is part of the PBC Study Group and received a positive review in March 2018~\cite{pbc}.


\end{document}